\documentstyle[emulateapj]{article}
\def\gsim{\;\rlap{\lower 2.5pt
 \hbox{$\sim$}}\raise 1.5pt\hbox{$>$}\;}
\def\lsim{\;\rlap{\lower 2.5pt
   \hbox{$\sim$}}\raise 1.5pt\hbox{$<$}\;}
\makeatletter
\newenvironment{tablehere}
  {\def\@captype{table}}
  {}
\newenvironment{figurehere}
  {\def\@captype{figure}}
  {}
\makeatother

\def\be{\begin{equation}}
\def\ee{\end{equation}}
\def\bea{\begin{eqnarray}}
\def\eea{\end{eqnarray}}

\def\cmm2{{\,\rm cm^{-2}}}
\def\cm2{{\,{\rm cm}^2}}
\def\cmm3{{\,{\rm cm}^{-3}}}
\def\gcmm3{{\,{\rm g\,cm^{-3}}}}

\def\la{\mathrel{\mathpalette\fun <}}
\def\ga{\mathrel{\mathpalette\fun >}}
\def\fun#1#2{\lower3.6pt\vbox{\baselineskip0pt\lineskip.9pt
  \ialign{$\mathsurround=0pt#1\hfil##\hfil$\crcr#2\crcr\sim\crcr}}}

\lefthead{HAIMAN \& KNOX}
\righthead{FIRB CORRELATIONS}

\begin{document}\begin{flushright}
{\footnotesize
FERMILAB-Pub-99/189-A}
\end{flushright}
\nopagebreak
\vspace{-\baselineskip}
\submitted{Accepted for publication in ApJ}
\title{Correlations in the Far Infrared Background}
\author{Zoltan Haiman}
\affil{NASA/Fermilab Astrophysics Center\\ Fermi National Accelerator
Laboratory, Batavia, IL 60510, USA, email: zoltan@fnal.gov}
\and
\author{Lloyd Knox}
\affil{Department of Astronomy and Astrophysics,
University of Chicago, Chicago, IL 60637, USA, email: knox@flight.uchicago.edu}

\begin{abstract}
We compute the expected angular power spectrum of the cosmic Far Infrared
Background (FIRB).  We find that the signal due to source correlations
dominates the shot--noise for $\ell \la 1000$ and results in anisotropies with
rms amplitudes $(\sqrt{\ell(\ell+1)C_\ell/2\pi})$ between 5\% and 10\% of the
mean for $l \ga 150$.  The angular power spectrum depends on several unknown
quantities, such as the UV flux density evolution, optical properties of the
dust, biasing of the sources of the FIRB, and cosmological parameters.
However, when we require our models to reproduce the observed DC level of the
FIRB, we find that the anisotropy is at least a few percent in all cases. This
anisotropy is detectable with proposed instruments, and its measurement will
provide strong constraints on models of galaxy evolution and large-scale
structure at redshifts up to at least $z \sim5$.
\end{abstract}

\keywords{cosmology: theory -- cosmology: observation -- cosmology: far
infrared background -- cosmic microwave background -- galaxies: formation --
galaxies: evolution}

\section{Introduction}

The recent discovery of the cosmic Far Infrared Background (FIRB,
\cite{puget96,fixsen98,dwek98,schlegel98,lagache99}) and the determination of
its spectrum creates the opportunity for a new probe of structure formation in
the high--redshift universe (e.g. \cite{guiderdoni98,blain99a}).  One way of
improving our understanding of this background is to observe the sky with deep
exposures at high angular resolution, to determine the number counts of
discrete sources
(\cite{hughes98,barger98,eales99,smail97,holland98,blain99b,barger99,puget99}),
and the redshift distribution of the contributing sources
(e.g. \cite{barger99b,blain99c}).  Observations with the Sub--millimetre Common
User Bolometer Array (SCUBA) camera (\cite{holland98}) have indeed identified
point sources at 450$\mu$m and $850\mu$m, which account for a large fraction
($\gsim 25\%$) of the FIRB at these wavelengths.  However, a conclusive
identification of optical counterparts, or the determination of the redshifts
of these sources are still made difficult by the lack of detailed spectral
information, and the uncertainty in the SCUBA-determined positions.

A complimentary approach is to observe the background at relatively low angular
resolution in the ``confusion limit'' with the aim of studying its statistical
properties.  Such analyses have proven fruitful in the context of the Cosmic
Microwave Background (CMB), as well as for the optical (\cite{vogeley97}) and
near--infrared (\cite{JimKas97}, \cite{kas97}) backgrounds.  The aim of the
present {\it Letter} is to predict the lowest--order statistical property of
the FIRB, the two--point angular correlation function, using simplified models
for the origin of the background flux.  Our results demonstrate that under
conservative assumptions, the correlations are measurable with {\it
Planck}\footnote{{\it Planck}:
http://astro.estec.esa.nl/SA-general/Projects/Planck/} and other future
instruments, such as the balloon--borne {\it Far Infrared Background Anisotropy
Telescope} ({\it FIRBAT}) proposed by the TopHat\footnote{{\it TopHat}:
http://topweb.gsfc.nasa.gov} group and the {\it Bolometric Large Aperture
Sub-mm Explorer} ({\it BLASE}, \cite{blase}).  In addition, the signal might be
detectable in the highest frequency channel of the existing {\it
BOOMERanG}\footnote{{\it BOOMERanG}: http://www.physics.ucsb.edu/\~boomerang}
data.

To calculate the correlation function, we must first understand the nature of
the source(s) of the background.  A compelling explanation is thermal emission
by the interstellar dust of high--redshift galaxies, heated by their internal
optical and ultraviolet (UV) star--light.  A handful ($\sim 30$) of sources
with the expected properties have been resolved in several SCUBA images
at $850\mu$m, reported in several papers
(\cite{hughes98,barger98,eales99,smail97,holland98,barger99,blain99b}) and
summarized by Scott \& White (1999), and another handful at 
$175\mu$m with the ISOPHOT instrument on board the 
Infrared Space Observatory (ISO) (\cite{puget99}).  
According to Hughes et al. (1998) and
Barger et al. (1999), sources brighter than 2mJy comprise at least 25\% of the
background at 850$\mu$m (350 GHz).  The observed source counts can be
parameterized empirically by a softened power-law form (e.g. \cite{barger99}).
Using this form to extrapolate towards the faint end of the source count
distribution (down to $\sim0.5$mJy), one can account for the entire background.
These conclusions are further supported by the somewhat less reliable
detections of faint sources down to $\sim$1mJy.  One should bear in mind,
however, that such extrapolations are highly sensitive to the properties of the
high--z population.

The sources identified by SCUBA are possibly galaxies.  Two have been
conclusively identified as such (\cite{frayer98}).  Hughes et al. (1998),
observing in the Hubble Deep Field (HDF), identified tentative optical
star--burst galaxy counterparts for all five of their sources brighter than
2mJy; four of them with redshifts between 3 and 4 and one with $z \simeq 0.9$.
These identifications are not yet conclusive, due to the coarse angular
resolution of SCUBA, and the possibility remains that many of the sources are
dust-enshrouded quasars, rather than galaxies (see, e.g. \cite{sanders99}).

It is also not yet clear that the FIRB is entirely due to the type of sources
seen with SCUBA.  Other possibly significant contributors include emission from
inter-galactic dust, ejected from galaxies by radiation pressure or supernova
winds (\cite{aguirre99,ah99}), numerous low surface brightness, dusty
protogalaxies, and, more speculatively, radiatively-decaying massive particles
(\cite{bch86}).  These contributions to the background will not be seen by
experiments that are targeting relatively bright point sources.  They would
all, however, contribute not only to the mean level of the background, but also
to its fluctuations on large angular scales, as long as their infrared emission
traces the spatial mass fluctuations to some degree.

In this paper, we do not model the galaxy evolution process in any detail, but
instead adopt a toy model that allows us to explore the dependence of the FIRB
angular power spectrum on the nature of the sources, their redshift
distribution, and cosmological parameters.  More sophisticated models exist in
the literature that generalize semi--analytical galaxy evolution schemes to
make detailed predictions in the infrared regime (e.g.
\cite{toffolatti98,guiderdoni98,blain99a}).  However, even these models are
very much driven by the data and the semi-analytic approach has resulted in a
wide range of predictions in the past.  Here we choose a simplified approach,
which is sufficient to illustrate the detectability of the clustering signals,
their dependence on the large-scale distribution of matter and aspects of
galaxy formation, and to demonstrate the need for further work.

An approach even simpler than our own is that of Scott and White (1996) who
estimated the angular correlation function of the FIRB at 850$\mu m$ by
assuming that the FIRB sources have an angular correlation function like that
of Lyman--break galaxies (\cite{giavalisco98}).  They viewed the FIRB as a
contaminant of the CMB data, and were interested in a rough estimate of what
these correlations might possibly be.  As a result, they could ignore the fact
that the projection effects would be different for the two different classes of
objects.  In the present paper, we investigate the FIRB as an interesting
signal in its own right, and model it in sufficient detail to understand the
dependence on large-scale structure and galaxy formation.

The spectra and correlation functions of both the near and far infrared
backgrounds have been considered in general terms in the pioneering works of
Bond, Carr \& Hogan (1986, 1991).  The recent measurements of the FIRB, the
discrete source detections discussed above, as well as determinations of the
redshift--evolution of the global average star formation rate (e.g.
\cite{madau99} [M99]), the UV background at $z=0$ (\cite{bernstein97}), and at
$2\lsim z \lsim 4$ (\cite{prox96}) now allow a more focused discussion.  When
we calibrate our models using the recent infrared data, we find that the
clustering results in contrasts of about 10\% in the FIRB, which is
sufficiently strong to dominate the shot--noise (estimated from the SCUBA
detections) and to be detectable by proposed future missions.  \cite{JimKas97}
have calculated the angular power spectra for the {\it near} infrared
background, and also find contrasts of roughly 10\%.  Indeed, very recently,
\cite{kas99} have claimed a tentative detection of fluctuations of cosmic
origin in the DIRBE (Diffuse Infrared Background Explorer) maps 
from 1 to 5 microns, with the expected amplitude,
or possibly slightly higher.  The angular power
spectrum in Scott and White (1996) is bracketed by those in our range of
models, although most of our models tend to have somewhat higher amplitudes.

Our work has been strongly motivated by future missions including {\it Planck}
and the {\it FIRBAT}.  Both are capable of detecting these fluctuations, the
{\it FIRBAT} by observing in eight channels with central wavelengths ranging
from 230$\mu$m to 940$\mu$m with per--channel sensitivities between 25 and 130
$\mu{\rm K~s^{1/2}}$, and an angular resolution of $6'$.  The broad range of
frequencies is important for separating the FIRB fluctuations from those of
dust in our own galaxy.  It is possible that this signal is discernible in the
existing {\it BOOMERanG} data, although the limited spectral coverage will make
discrimination from dust in our own galaxy difficult.

The rest of this {\it Letter} is organized as follows.  In \S~2, we describe
our toy model for the mean FIRB.  In \S~3, we extend this model to include the
fluctuations, by assuming cold dark matter (CDM) power spectra, and that the
FIRB light is a (biased) tracer of mass.  In \S~4, we show results in several
models, including those with a uniform bias, and a redshift--dependent bias
calculated according to a prescription for galactic halos (\cite{mowhite96}).
In \S~5, we discuss the results, and what can be learned from planned
observations of the FIRB on large ($\ga 5'$) angular scales, and finally in
\S~6, we summarize our conclusions.

\section{Modeling the Mean FIRB}

In our model, the FIRB arises from thermal dust emission.  To compute the mean
level of the flux, the main ingredients are the mass density $\Omega_{\rm d}$,
and temperature $T_{\rm d}$ of dust, and the evolution of these quantities with
redshift. The angular fluctuations depend further on the spatial distribution
of the dust, as we will discuss in \S~3 below.  In order to compute the
evolution of $\Omega_{\rm d}(z)$ and $T_{\rm d}(z)$, we rely on the recent
determinations of the evolution of the global average star--formation rate
(SFR, see \cite{lilly96}, M99 and references therein).  In particular, we
assume that both the UV emissivity that determines the dust temperature, and
the rate of dust production, are proportional to the global SFR (corrected for
dust absorption), $\dot \rho_\star(z)$, as given by M99.  Our motivation for
relying on the star--formation history is that the stellar UV flux is known to
dominate that of quasars at all redshifts $z\lsim 5$, with quasar contributions
less than $\sim20\%$ (M99).  We assume further that the dust has a composition
of 50\% graphites and 50\% silicates by mass, and composite cross--sections as
in the Milky Way (\cite{drainelee84}). Finally, we assume that the UV
emissivity has the spectral shape obtained by summing individual zero--age main
sequence stellar spectra weighted by a Scalo (1986) mass function
(\cite{bc96}).

\begin{tablehere}[t]
\caption{\label{tab:models} \footnotesize The parameter values in our standard
model and its variants.  The last two columns are the values of the dust
density and temperatures at $z=0$ (in Kelvin), determined from fitting the FIRB
at 850$\mu$m.  See text for discussion.}
\vspace{0.3cm}
%\begin{center}
\noindent\begin{tabular}{|c||c|c|c|c|c||c|c}
\hline
   & $\Omega_{\rm m}$   & $\Omega_\Lambda$ & $h$ & SFR & $T_{\rm Gr}/T_{\rm Si} $ & $\Omega_{\rm d}$ \\
\hline
\hline  
Stand.  & 0.3 & 0.7 & 0.65 & M99   & 14.4/10.1 & $2.7\times10^{-5}$ \\
Open    & 0.3 & 0.0 & 0.65 & M99   & 13.9/9.8  & $3.8\times10^{-5}$ \\
SCDM    & 1.0 & 0.0 & 0.5  & M99   & 14.1/9.9  & $5.4\times10^{-5}$ \\
Flat    & 0.3 & 0.7 & 0.65 & flat  & 17.1/12.9 & $3.5\times10^{-5}$ \\
Hot     & 0.3 & 0.7 & 0.65 & BSIK  & 40.0/40.0 & $2.0\times10^{-6}$ \\
High-z  & 0.3 & 0.7 & 0.65 & $z=7$ & 12.9/9.1  & $2.9\times10^{-5}$ \\
\hline
\end{tabular}
\vspace{0.3cm}
\end{tablehere}

Under these assumptions, the dust temperature is determined by the UV
radiation, i.e. by requiring that the amount of energy absorbed in the UV is
equal to that thermally radiated in the infrared.  The proportionality
constants for the dust production rate and the UV emissivity are then
determined by fitting the observed spectrum of the mean FIRB.  The relevant
equations are summarized in \S~3 of Aguirre \& Haiman (1999) (see also
\cite{lh97}).  Note that we have not assumed that the dust is perfectly
uniformly distributed, but only that wherever there is dust, it is exposed to
the same radiation field.  Also note that in this scenario, as long as we are
interested only in the mean level of the FIRB, we do not need to specify the
spatial distribution of the dust, or the type of source from which the infrared
emissivity arises.  To get the angular power spectrum, we will assume below
that dust is a (biased) tracer of mass on large scales (see \S~3).

The parameters of our models are summarized in Table~\ref{tab:models}.  Our
standard model is a $\Lambda$CDM cosmology, with the SFR taken from M99.  In
this model, we find that in order to fit the FIRB, the graphite/silicate dust
temperatures at $z=0$ need to reach 14.4K and 10.1K, while the total mass
density of dust has to reach $\Omega_{\rm d}=2.7\times10^{-5}$.  The redshift
evolution of these quantities are shown in the upper panel of
Figure~\ref{fig:emiss}.  In particular, $\Omega_{\rm d}$ rises continuously as
dust is accumulated from the increasing number of stars.  The dust temperatures
deviate from the CMB at $z\approx 7$, and stay roughly constant (silicates at
$\sim$15K, graphites at $\sim$25K) until $z \sim 1$, at which point the SFR
drops sharply, the dust heating becomes less efficient, and the dust
temperatures start dropping again.  The dust temperatures we obtain in our
fiducial model at $z=0$ are somewhat colder than, e.g. the temperature of dust
in the Milky Way ($\sim 16-20$K, \cite{reach95}).  The temperatures are higher
in our ``flat'' SFR model ($\sim 17/13$K, see Table~\ref{tab:models} and
discussion below), and are in better agreement with the Milky Way value.  We
therefore note that our conclusions are not dependent on an accurate
reproduction of dust temperatures in local galaxies.  In the lower panel of
Figure~\ref{fig:emiss}, we show the redshift evolution of the comoving dust
emissivity $j_{\rm d}(z)=\rho_{\rm d}\times \kappa_{{\rm d},\nu(1+z)}\times
B_{\nu(1+z)}[T_{\rm d}/(1+z)]$ in units of $10^{-21}~{\rm
erg~cm^{-3}~s^{-1}~sr^{-1}~Hz^{-1}}$ at the (observed) wavelengths of 850$\mu$m
(lower solid curve) and 450$\mu$m (lower dashed curve).  Here $\kappa$ is the
dust opacity coefficient and $B_\nu$ is the Planck function.  The curves in the
figure illustrate the faster fall--off towards higher redshifts of the
emissivity at $450\mu$m compared to $850\mu$m, due to the fact that the peak of
the grey body dust {\it observed} at 450$\mu$m occurs at lower $z$ than for the
longer wavelength $850\mu$m.

%########################## figure 1 ########################################
\begin{figurehere}
\epsscale{1.0}
\plotone{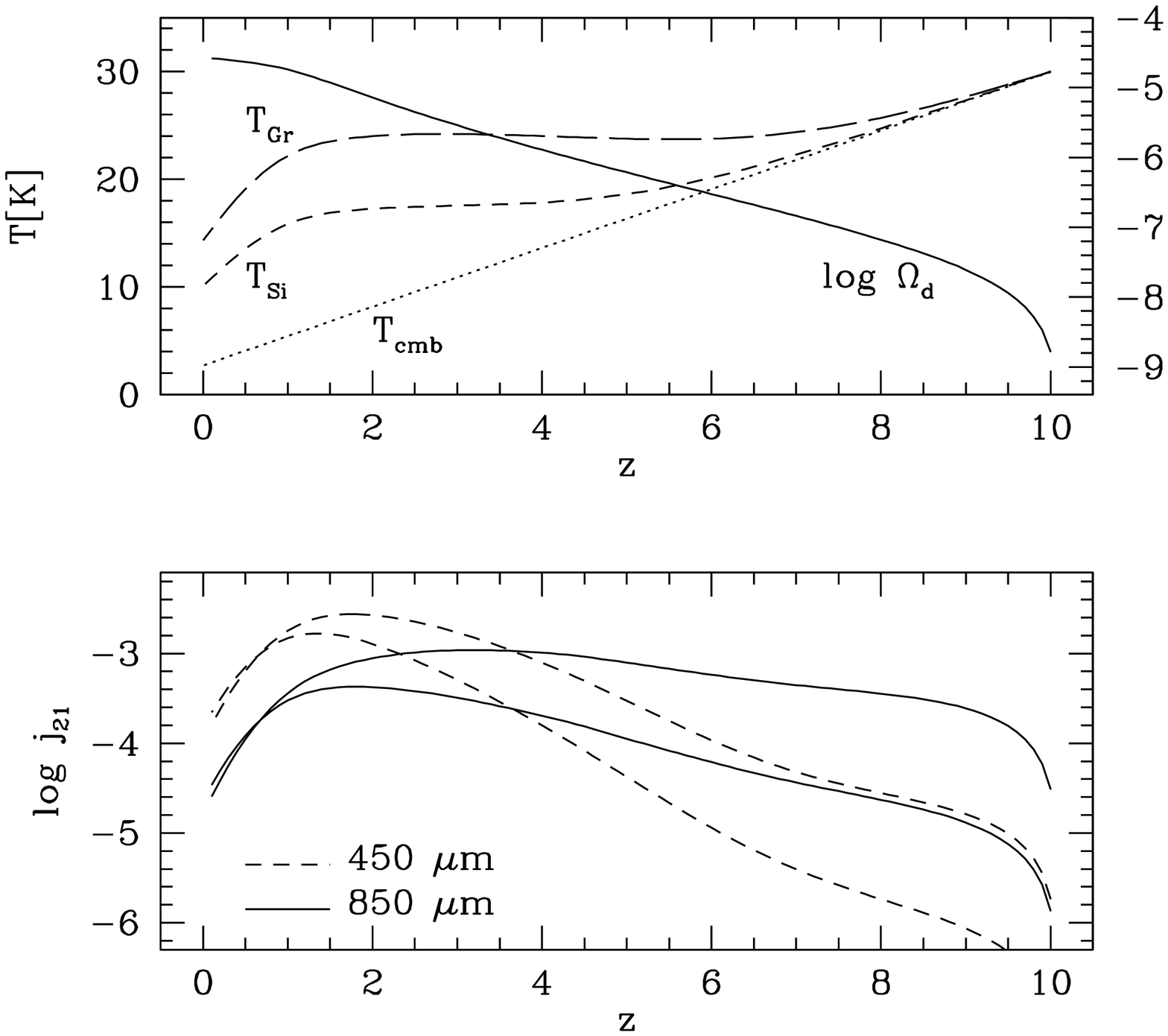}\vspace{-0.4cm}
\caption[]{\label{fig:emiss} Upper panel shows the evolution of dust
temperatures in our standard model, compared with the CMB temperature, as well
as the mass density of dust.  The labels on the right axis refer to
$\log\Omega_{\rm d}$.  Lower panel shows the comoving dust emissivity $j_{\rm
d}$ at 450$\mu$m and 850$\mu$m (bottom curves).  The top curves show the
quantity $bj_{\rm d}$, where $b$ is the bias parameter of galaxy sized
($10^{12}~{\rm M_\odot}$) halos. }
\end{figurehere}
\vspace{0.3cm}
%############################################################################

We considered several variations on our standard model, summarized in
Table~\ref{tab:models}.  To explore the dependence of our results on the
underlying cosmology, we have evaluated an open CDM model with $\Omega_{\rm
m}=0.3$, and a standard CDM model.  We then studied a ``flat'' SFR model, in
which the SFR is constant in redshift. The motivation for considering
this model is that the UV flux that heats the dust inside an individual
galaxy may not evolve in the same way as the global average SFR does.  Indeed,
the evolution of the global SFR is driven mainly by the change in the number of
galaxies, rather than the evolution of the characteristic brightness
(\cite{madau97}).  In this case, the dust temperature within each galaxy might
stay roughly constant, provided it is heated by long--lived stars.  To mimic
this scenario, we have included a decreasing overall dust abundance according
to the original M99 SFR, but set the UV emissivity to be a constant at all
redshifts.   

As an alternative to this scenario, we also considered a ``hot'' dust model,
adopted from \cite{blain99c}.  In this model, the dust in each galaxy is kept
hot by the more intense radiation from short--lived, intense bursts of
star--formation, after which the dust cools down, and its emission is
subsequently ignored.  The total dust emission at a given redshift in this case
is proportional to the UV emissivity, which, in turn, is proportional to the
SFR.  Following \cite{blain99c}, we assume a constant dust temperature of
$40$K, and a star--formation rate that is, in effect, a sum of the M99 SFR, and
an additional Gaussian peak centered on redshift $z=2.1$.  Note that in this
model, only a fraction of the total dust abundance is ``luminous'' at any given
time -- the entry for $\Omega_{\rm d}$ in Table~\ref{tab:models} refers to the
maximum luminous dust density, reached at $z=2.1$.  Finally, we consider a
``high--$z$'' variation on our standard model, in which we postulate an
additional peak in the SFR at redshift $z=7$, with SFR
$\propto\exp[-(z-7)^2/2]$, and an amplitude such that this hypothetical
high--redshift population accounts for 50\% of the FIRB at $850\mu$m.  The
choice of this large burst of star formation is somewhat ad--hoc.  Our
motivation for considering this scenario is to characterize the possibility
that the FIRB has a significant contribution from very high redshifts -- an
option that can not be ruled out by present observations.  Note that
significant star--formation at redshifts $z\gsim 5$ can still have escaped
detection by, e.g., HST.

%########################## figure 2 ########################################
\begin{figurehere}
\epsscale{1.0} 
\plotone{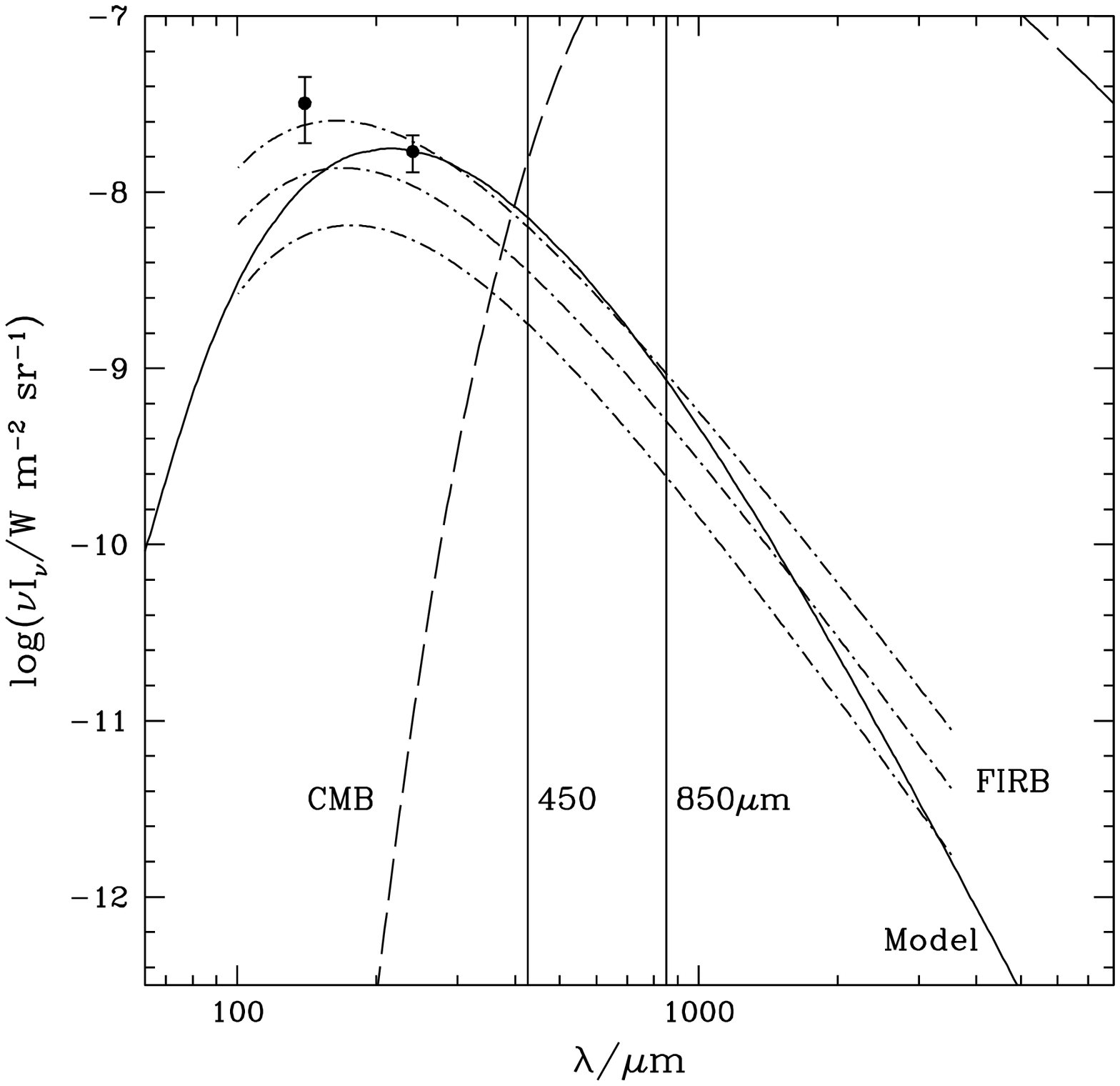}\vspace{-0.4cm}
\caption[]{\label{fig:spec} The spectrum of the FIRB predicted in our standard
model (solid curve).  The dot--dashed curves show the measurements with
$\pm1\sigma$ uncertainties from Fixsen et al. (1998), and the long dashed
curves show the CMB for reference.  The data--points are taken from Schlegel et
al. 1998.}
\end{figurehere}
\vspace{0.3cm}
%############################################################################

In Figure~\ref{fig:spec} we show the spectrum of the FIRB in our standard
model.  The dot--dashed curves show the fit to the measurements by Fixsen et
al. (1998), while the solid line shows our model prediction.  The long--dashed
lines show the CMB, and the vertical solid lines mark the observational
wavelengths of 450$\mu$m and 850$\mu$m. As the figure reveals, the prediction
from our model is in reasonably good agreement with the FIRB; in particular, it
is within the quoted $\pm 1\sigma$ error bars.  The quality of fits in the
other models from Table~\ref{tab:models} is similar to that of the standard
model shown in Figure~\ref{fig:spec}.  The real sources of the FIRB are much
more complex than our simplified model assumes.  The UV flux density and the
dust density are not expected to be spatially uniform, and dust in different
galaxies could have different temperatures, broadening our predicted
dust--emission peak.  However, our model is sufficient to allow for a rough
prediction for the angular power spectra at various wavelengths, and to provide
the framework for a discussion of what might happen with more realistic models.

\section{FIRB Fluctuations}

The model described so far accounts for the amplitude of the FIRB, but makes no
specific reference to the spatial distribution of the infrared--emitting dust.
In this section, we characterize the spatial distribution in order to derive
the angular power spectrum of the FIRB.  One simple assumption is that ``dust
traces mass'', i.e. that the local density of dust is proportional to the total
mass density at every point in the universe.  In reality, most of the dust
could be confined in galactic halos; in which case the spatial distribution of
dust emission would follow that of galactic halos.  The correlation function of
dark halos is related (\cite{mowhite96}) to that of mass by the bias parameter
$b=b(M_{\rm halo},z)$. Here we adopt the modified formula of Jing (1999), which
has been shown to reproduce the results of numerical simulations on a wider
range of scales (including $M_{\rm halo}<M^*$). Under these assumptions, the
bias $b$ is scale--independent for a fixed halo size, but evolves in
redshift. The correlation function between pixels separated by angle $\theta$
and at frequencies $\nu$ and $\nu'$ is given by: \bea C^{\nu \nu'}(\theta) &=&
\sum_l{2l+1\over 4\pi}C_l^{\nu \nu'} P_l(\cos\theta) \\ C_l^{\nu \nu'} &=&
{2\over \pi} \int k^3 P(k) f_{\nu l}(k) f_{\nu' l}(k) {dk\over k} \\ f_{\nu
l}(k) &\equiv& \int j_l(kr)b(r)D(r)j_d(\nu,r)a(r)dr \eea where $r \equiv
c(\eta_0 - \eta)$ is the coordinate distance of an event at conformal time
$\eta$ on our past light cone, $\eta_0$ is the conformal time today, $P(k)$ is
the power spectrum of the matter today, $a(r)$ is the scale factor normalized
so that $a(0)=1$, $D(r)$ is the linear theory growth factor, and $j_l$ is the
spherical Bessel function. In a matter--dominated universe, $D(r) = 1/a$.  For
the more general case of non--zero curvature and/or non--zero cosmological
constant we use the fitting formula of Peebles (1980) and Caroll, Press \&
Turner (1992), respectively.

Equations 2 and 3 are a version of Limber's equation
(\cite{limber53,peebles80}), although with power spectra instead of correlation
functions and generalized to describe correlations between different
components.  Note that pixels at unequal frequencies will not be perfectly
correlated because $j_d(\nu,z)$ and $j_d(\nu',z)$ are not proportional to each
other for $\nu \ne \nu'$.  The unequal--frequency correlation function (at
$\theta=0^\circ$, with $5'$ smoothing) has been calculated by Bouchet \&
Gispert (1999) by using simulated Planck maps generated using the
semi--analytic model of Guiderdoni et al. (1998).

As equations~(1--3) show, $C_l$ scales roughly as $b^2$.  To illustrate the
effect of a non--negligible bias, in the lower panel of Figure~\ref{fig:emiss}
we show the evolution of the comoving emissivity $j_d(\nu,z)$ at 450$\mu$m and
850$\mu$m (bottom curves), together with the quantity $j_d(\nu,z)b(M_{\rm
halo},z)$ (top curves).  We have assumed $M_{\rm halo}=10^{12}~{\rm M_\odot}$,
which is valid if dust emission arises from dark halos of this size (roughly
the size of galactic halos, as well as the host halos of typical quasars, see
\cite{hl98}).  The figure shows that biasing significantly boosts the
contribution to the signal from $z\gsim 1$, but has negligible effect at
$z\lsim 1$.  In what follows, we alternately use this prescription for biasing
and a time--independent bias of $b=3$, which is roughly that of Lyman--break
galaxies at $z \sim 3$ (\cite{giavalisco98}).

%########################## figure 3 ########################################
\begin{figurehere}
\epsscale{1.0} 
\plotone{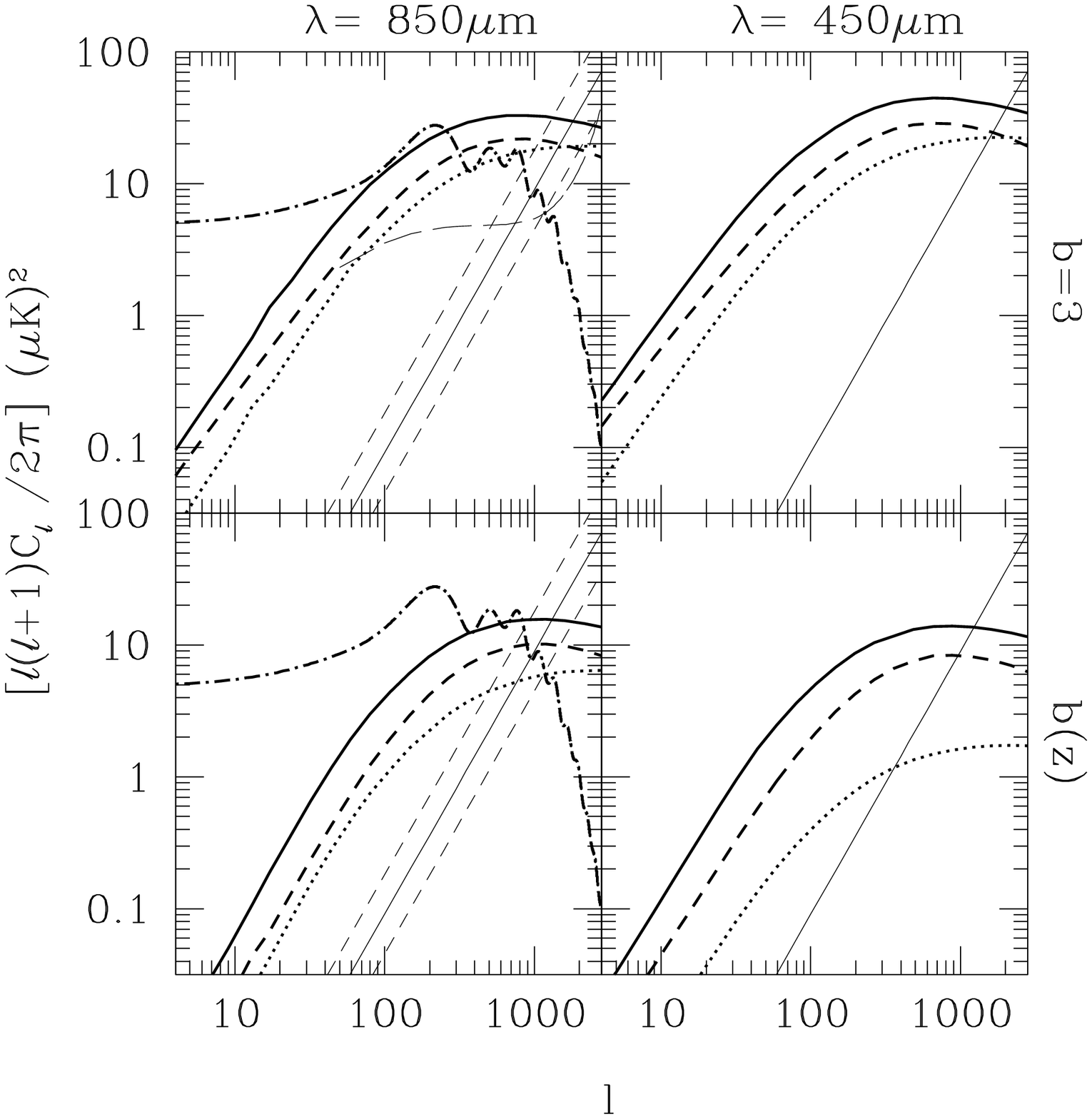}\vspace{-0.4cm}
\caption[]{\label{fig:cls1} Predicted angular power spectra in units of
$(\mu{\rm K})^2$ (antenna temperature) for different cosmologies: $\Lambda$CDM
(solid), SCDM (dashed) and OCDM (dotted), all assuming M99 SFR.  The upper
panels assume linear bias with a constant $b=3$, and the lower panels assume a
$z$--dependent bias.  The left panels show the results at 450$\mu$m and the
right panels at 850$\mu$m.  The long-dashed line on the upper left panel shows
the sensitivity of {\it FIRBAT}, assuming bins of $\delta l = 50$ and a one
week flight.  For reference, the CMB power spectrum is also shown (dot--dashed
lines in the 450$\mu$m panels).  The light solid lines (with bounding dashes
for the $450\mu$m panels) are estimates of the shot--noise power with
indications of the uncertainty.  We do not attempt to characterize the
uncertainty at $\lambda = 450\mu$m.}
\end{figurehere}
\vspace{0.3cm}
%############################################################################

For our standard ($\Lambda$CDM) model, we have adopted the fitting formulae for
the power spectrum given by Eisenstein \& Hu (1999) with power--law index $n=1$
and normalization $\sigma_8=1$, while in the Open and SCDM models, we used
($n=1.3$, $\sigma_8=0.85$) and ($n=0.7$, $\sigma_8=0.6$), respectively.  These
values for $n$ and $\sigma_8$ were chosen to roughly agree with both the
cluster abundance (\cite{VL99}) and COBE--DMR (\cite{BunnWhite96}) constraints.
For the conversion between coordinate distance and scale factor for the
$\Lambda$CDM model, we used the fitting formula of Pen (1999).

The total $C_l$ is the sum of that due to correlations, which we have
calculated, and that due to the discrete nature of the sources.  This
``shot--noise'' $C_l$ was calculated by Scott \& White (1999) at $\lambda = 850
\mu$m who used the double--power law LF constrained by SCUBA data mentioned
earlier and is shown in Fig.~3 rising like $l^2$ (as all white--noise power
spectra do).  The LF at $\lambda = 450 \mu$m is much more uncertain; we
estimate the shot-noise at this wavelength simply by assuming that the {\it
shape} of the LF is independent of frequency.  Since the antenna temperature of
the mean FIRB is roughly the same at these two wavelengths ($T_{\rm mean}
\simeq 60 \mu$K), this means that the shot--noise at 450$\mu$m is the same as
at 850$\mu$m.

\section{Results}

Our results are plotted in Figures~\ref{fig:cls1} and \ref{fig:cls2}.
Figure~\ref{fig:cls1} shows the predicted angular power spectra in units of
$(\mu{\rm K})^2$ (antenna temperature) for different cosmologies: $\Lambda$CDM
(solid), SCDM (dashed) and OCDM (dotted), all assuming M99 SFR.  The upper show
our models with a constant linear bias $b=3$, and the lower panels show models
with a $z$--dependent bias.  The left panels show the results at 450$\mu$m and
the right panels at 850$\mu$m.  The long-dashed line on the upper left panel
shows the sensitivity of {\it FIRBAT}, assuming bins of $\delta l = 50$ and a
flight duration of $6\times10^5$s.  For reference, the CMB power spectrum is
also shown (dot--dashed lines) in the 850$\mu$m panels.  The light solid lines
with bounding dashes in the same panels are estimates of the shot--noise power
with indications of the uncertainty.  We do not attempt to characterize the
uncertainty at $\lambda = 450\mu$m.  Figure~\ref{fig:cls2} demonstrates how our
results change due to variations away from our standard model (solid line): we
show the flat SFR model (dotted), hot dust model (dashed), and high--$z$ peak
model (dot--dashed).

%########################## figure 4 ########################################
\begin{figurehere} 
\epsscale{1.0} 
\plotone{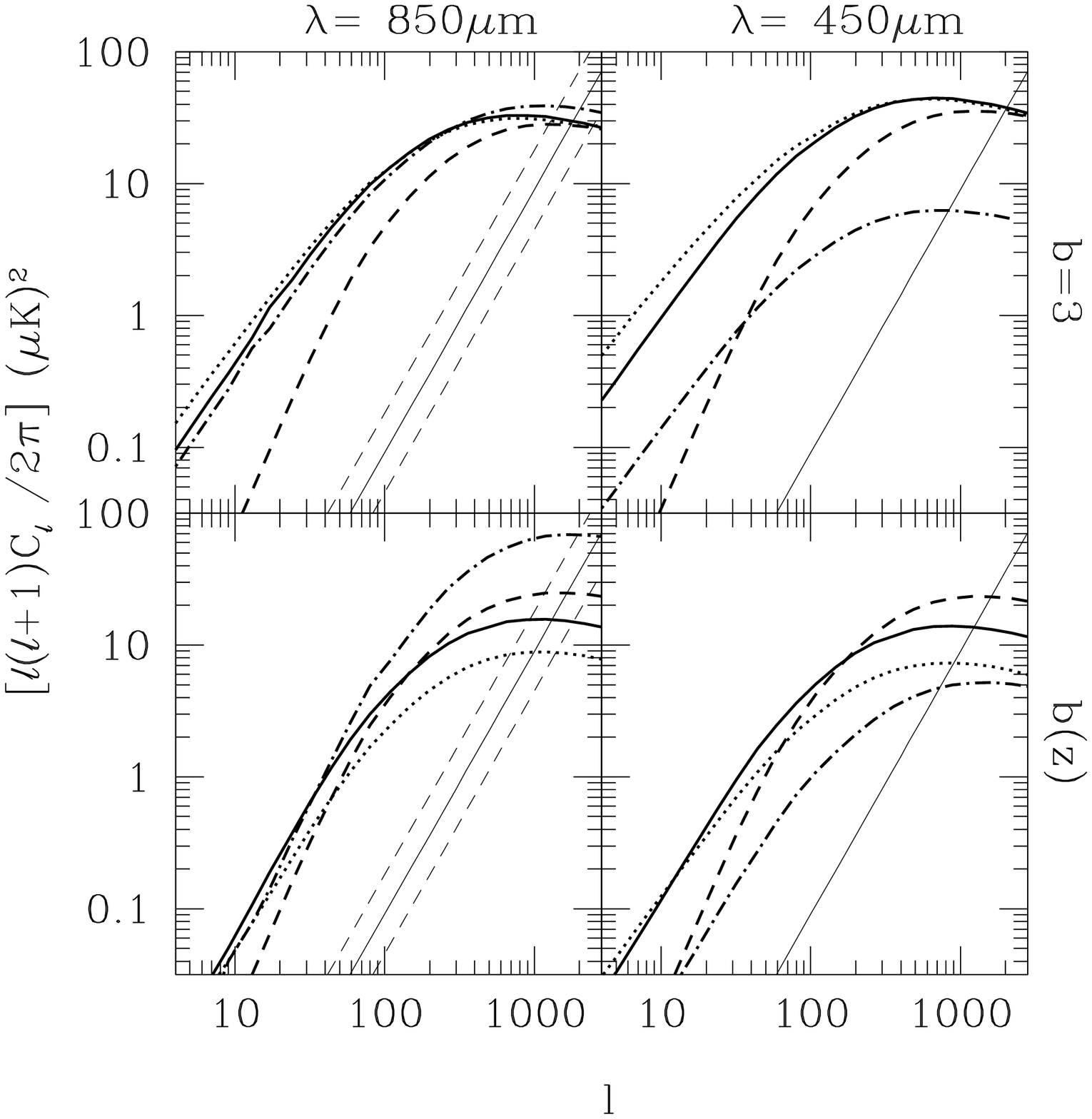}\vspace{-0.4cm}
\caption[]{\label{fig:cls2} Angular power spectra as in Figure~\ref{fig:cls1},
but showing variations away from our standard model (solid line):
flat SFR (dotted), hot dust (dashed), and high--$z$ peak in SFR (dot--dashed).}
\end{figurehere}
\vspace{0.3cm}
%############################################################################

The first thing to note is that the contrast
($\sqrt{l(l+1)C_l/(2\pi)}$) between $\ell \simeq 150$ and $\simeq 1000$ is
about 10\% of the $\tilde 60\mu$K mean and is mostly due to the 
intrinsic clustering of the
FIRB, rather than shot--noise.  We can roughly understand this basic result
analytically with the following expression: 
\be 
{l(l+1) C_l \over 2\pi} \sim
T_{\rm mean}^2 \left[{b \over (1+z_{\rm peak})}\right]^2 \Delta^2(k)
{\pi/k\over \sigma} 
\ee 
where $\Delta^2(k)\equiv k^3 P(k)/(2\pi^2)$ is the contribution to the variance
of the matter density from each logarithmic interval in $k$, $\sigma$ is the
coordinate distance over which there is substantial emission, and $k \simeq
l/r_{\rm peak}$ where $r_{\rm peak}$ and $z_{\rm peak}$ are the coordinate
distance and redshift of the peak emission.  The right--most term is the
inverse square of the ``wash--out factor'' arising from the incoherent summing
of structure at different redshifts, which reduces the contrast.  Taking
$z_{\rm peak} \simeq 1$ and $l=1100$ we find $k\simeq 0.5 h Mpc^{-1}$ 
and $\Delta^2(k) = 3$
for our standard model.  Further taking $b=3$ and $\sigma \simeq 3500$
h$^{-1}$Mpc we find $[{l(l+1) C_l \over 2\pi}]_{l=1100} \simeq 40 \mu$K$^2$.
The shape of $l(l+1)C_l/(2\pi)$ (rising from low $l$, plateauing, and then
slowly dropping) is due to a) the fact that $\Delta^2(k) \propto k^{3+n}$ at
low $k$, flattening out to $k^{n-1}$ at high $k$ and b) the wash--out factor
resulting in another factor of $k^{-1}$.

Note that the only quantities that matter for $C_l$ are the matter power
spectrum, $a(r)$, and the product $b(r)D(r)j_d(r)$.  With the matter power
spectrum and $a(r)$ fixed, any effect that pushes the $bDj$ product to peak at
higher redshifts, shifts the spectrum to higher $l$.  Thus, for the constant
bias cases, the flat SFR model
has slightly more power at low $l$ and the high--$z$ SFR model is shifted to
higher $l$.  Likewise, the hot dust model has the steepest rise
from low $l$ to high $l$ because its emissivity at both 450$\mu$m and 850$\mu$m
rises from low $z$ more quickly than any of the other models.

The redshift--dependent bias, $b(z)$, increases monotonically and at $z=$0, 2
and 7 has values of 1, 3 and 18 respectively.  Because $b(z<2) < 3$ the
redshift-dependent bias generally decreases the fluctuation power compared to
the $b=3$, $z$-{\it independent} case.  This is especially true at low $l$ and
also for $\lambda = 450\mu$m.  The exceptions are the high--$z$ SFR case and
the hot dust case at $\lambda = 850\mu$m, due to the 
significant emission from $z>2$.

\section{Discussion}

Our simplified model of the sources is insufficient for many purposes, e.g.,
predicting the luminosity function, but is adequate for predicting the
large--angular scale power to within factors of $\sim 2$.  The key assumption
is that the dust--light is a biased tracer of the mass; i.e., where there is
more mass there is proportionately more light.  Note that in principle, a
non--negligible fraction of the unresolved FIRB at 450$\mu$m, and especially at
850$\mu$m, could arise from the direct emission from optically faint quasars
(\cite{hl98}); our approach would be equally applicable in this case.  The
fluctuation level of 10\% is fairly model-independent, although linearly
dependent on the bias of the sources.  If the bias is unity, as is the case for
galactic halos near $z\approx0$ (\cite{mowhite96}), then the contrast may be as
small as a few per cent.

Contrasts observed to be smaller than a few percent would be surprising.  This
would indicate that most of the FIRB comes from sources with $b \la 1$, so that
$b(r)D(r)$ is small where $j(r)$ is large.  Examples of such sources would be
galaxies near $z\approx 0$, or intergalactic dust at high redshifts.  The
former case might be difficult to reconcile with the relatively low number of
nearby IRAS sources.  In the latter case, one would need to invoke dust
temperatures higher than expected based on the estimates of the UV background
(\cite{bernstein97}), in order to get the correct spectrum for the mean FIRB in
the full $150-1000\mu$m range (\cite{ah99}).

The correlated component can even be measured on angular scales where it is
smaller than the shot--noise.  The shot--noise contribution can be reduced by
observing at high resolution and masking out pixels with fluxes above some
threshold.  Fluctuation analyses of point--source cleaned SCUBA maps have been
done by two groups (\cite{hughes98,borys98}).  However, at these sub--arcminute
angular scales, the wash--out factor is very large, and the correlated
component is expected to be very small.  These analyses were motivated by the
desire to see the shot--noise due to unresolved sources below the flux cut, and
resulted in constraints on the faint--end slope of the number counts.

An actual detection of the FIRB fluctuations will be highly valuable.  The
spectrum of the FIRB {\it anisotropy} may eventually be better known than the
spectrum of the FIRB {\it mean} due to the experimental advantages of
differential measurements.  Such an improved determination of the FIRB spectrum
would provide detailed constraints on galaxy formation models, in addition to
those from the amplitude and scale--dependence of the anisotropy signal.  It
will also be interesting to examine the cross--correlations of the FIRB
fluctuations with other data such as the CMB (FIRB sources may lens the CMB,
or both backgrounds may be lensed by the same mass distributions), 
radio sources (from, e.g., the FIRST VLA survey or the {\it MAP} 
\footnote{{\it MAP}: http://map.gsfc.nasa.gov} 22 GHz channel), 
galaxies, or quasars at different redshift slices from the 2dF (e.g.,
\cite{folkes99}) and Sloan Digital Sky Survey (SDSS, see e.g., \cite{Gunn95}),
and the X--ray, and near infrared backgrounds 
(the same sources could contribute to
the backgrounds at both of these wavelengths).

Further development of predictions for the statistical properties of the FIRB
is clearly warranted.  The correlation function, the statistical property we
have focused on here, may prove easier to understand than the luminosity
function---the bright end of which is probably dominated by very rare events
(\cite{bond99}).  Correlation function predictions will be refined by detailed
modeling that is informed by additional observations.  One useful input will be
the clustering properties of galaxies and quasars, measured accurately in
forthcoming redshift surveys, such as 2dF and SDSS.  It may then be possible to
infer the nature of the sources of the FIRB by comparing its statistical
properties to those of galaxies and quasars.

In the preceding, we have implicitly assumed that the large-scale distribution
of matter ($P(k)$ and its evolution with redshift) will have been determined by
high-precision CMB anisotropy measurements and redshift surveys, and have
therefore been focusing on the dependence of $C_l$ on the nature of the
sources.  However, we emphasize that the FIRB $C_l$ is sensitive to large-scale
structure at redshifts intermediate to those that will be directly probed by
these redshift surveys ($z \la 1$) and CMB missions ($z \simeq 1100$).  It will
also be sensitive to wavelengths too small to be constrained by the CMB
measurements, and at any given comoving wavelength the matter fluctuations will
be better approximated by linear theory than they are at lower redshifts.  The
relation of $C_l$ to large-scale structure is complicated by its simultaneous
dependence on $j_d(\nu,z)$ and the bias properties of the sources,
but these complications will be reduced both by improved theoretical modeling
of the sources and also by high-resolution, deep observations, together with
identification of counterparts at other wavelengths.  Thus we view point source
observations and measurements of fluctuations on large angular scales as
complementary: the recovery of the power spectrum on large scales from $C_l$
observations is aided by measurement of the point sources, which is incapable
of determining the power spectrum alone.

\section{Conclusions}

The recently discovered cosmic Far Infrared Background and
the galaxy counts at 850$\mu m$ have opened a new
wavelength at which galaxy formation and evolution and large-scale structure
can be studied empirically.  A significant fraction of the FIRB has already
been resolved by SCUBA into discrete sources; however, due to the lack of
detailed spectral information, identified counterparts in other bands, and
secure redshift determinations (except for the two sources in
Frayer et al. 1998), fundamental questions still remain unanswered.
Does the population of the observed sources indeed account for the full
background? What is the nature of the observed sources: are they galaxies,
dust--enshrouded AGN, or a mixture of both?  What are their redshift
distributions?  We have argued that measurements of the FIRB correlation
function can help answer these questions and also that the FIRB provides a
unique probe of the large-scale distribution of matter at intermediate
redshifts.  We have found that under simple, but broad, models for the mean
FIRB, the anisotropy of the unresolved FIRB intensity is at the 3 to 10 percent
level.  These fluctuations are measurable with proposed balloon-borne
instruments and future space missions whose datasets will add new, useful
constraints on large-scale structure and models for the formation and evolution
of galaxies and quasars.

\acknowledgments 

We thank the TopHat group, and G. Wilson in particular, for asking what could
be learned from observing FIRB anisotropy, J.~R. Bond, S.~S. Meyer and
D.~N. Spergel for useful conversations, P. Madau for providing a fitting
formula for the SFR, and the referee, A. Blain, for a careful reading of the
manuscript and useful comments.  LK is supported by the DOE, NASA grant
NAG5-7986 and NSF grant OPP-8920223.  ZH was supported by the DOE and the NASA
grant NAG 5-7092 at Fermilab.

\end{document}